# Finite $Z$-less integral expressions for $\beta$-functions in the MS$^4$ scheme

N.D. Lenshina[1], A.A. Radionov[2], F.V. Tkachov[2]

[1] Physics Department, Lomonosov Moscow State University
[2] Institute for Nuclear Research, Russian Academy of Sciences, Moscow

lenshina.nd14@physics.msu.ru    alex.radionov89@gmail.com    fyodor.tkachov@gmail.com

**Abstract.** The generalized minimal subtraction scheme for ultraviolet renormalization (Kuznetsov and Tkachov, 1988) is fine-tuned with applications in mind. The resulting MS$^4$ scheme obviates extraneous regularizations and renders momentum integrands integrable by subtracting troublesome asymptotic terms in a physically correct fashion due to the use of special minimal subtraction operators defined congruously with the physically natural Polchinski cutoffs. A direct derivation of the Callan-Symanzik equations avoids divergent renormalization factors or counterterms, and automatically yields explicit exact finite integral expressions for renormalization group functions.

**1. Introduction.** Ultraviolet renormalization is a central feature of quantum field theory and deserves to be understood thoroughly and fully. But whether such an understanding has been achieved, is open to debate despite many cumbersome calculations successfully completed. This sentiment is supported by the findings we outline here after a preliminary report [1] and in advance of a detailed treatment [2].

The present work builds on a foundation laid in the 80's between refs. [3] and [4], specifically the theory of asymptotic operation (see sec. 2 below and refs. [5]–[7] for more details). The theory consists of distribution-theoretic techniques for handling products of singular functions and constructing their asymptotic expansions [8] and the family of generalized minimal subtraction (GMS) schemes for UV renormalization [9]. The former is a general mathematical basis for the latter which is a concrete physical formalism. Both grew organically out of the work on calculational problems with an understanding from the start that there can be no unfathomable magic behind the dimensional regularization (DR) [10] and the minimal subtraction (MS) scheme [11], although DR+MS did provide a fast track towards novel algorithms (see sec. 2) — even if it was not immediately apparent that computational efficiency may have been compromised thereby from the start (see the discussion in [1]).

Our purpose is to begin to explore the GMS formalism with applications in mind. Sec. 2 puts the work into context. Secs. 3–6 outline the MS$^4$ scheme as a special fine-tuning of the GMS scheme (the GMS family also includes the MS and $\overline{\text{MS}}$ schemes). With the scene set in sec. 3, sec. 4 defines a universal set of factorizable cutoffs whose shape is determined by a single function, essentially a Polchinski cutoff [12]. Sec. 5 uses the cutoffs to define a congruent set of the so-called special minimal subtraction operators that are the basic technical elements of the GMS/MS$^4$ framework. With such fine-tunings, in sec. 7, a per-graph version of the Callan-Symanzik equation is derived in a straightforward $Z$-less fashion, i.e. bypassing divergent renormalization factors, etc. Sec. 8 describes the so-called $\mathcal{B}$-operator that emerges from the derivation and that algorithmically maps an UV divergent graph to an explicit finite integral with all masses and external momenta set to zero and the scale invariance broken by the cutoffs. Sec. 9 outlines how (simply) renormalization group functions ($\beta$, $\gamma$ etc.) are assembled from multiloop 1PI graphs via the $\mathcal{B}$-operator. Sec. 10 discusses the findings, and sec. 11 draws conclusions.

**2. Motivations.** Given a byzantine history of the asymptotic operation (see [7] for a complete account), a few words are in order to put our work in context.

The DR+MS juggernaut opened floodgate for theorists' combinatorial ingenuity and hard work by providing easy-to-learn prescriptions to manipulate divergent integrals as if they were convergent, via a simple trick of keeping a symbolic dimensionality $D \neq 4$ as a regulator. At the end, the divergent parts are analytically extracted in the form of poles in $D-4$ and magically cancel. This manner of calculation enjoyed a massive practical success. Its visible feature is the ubiquity of renormalization factors $Z_i$ and counterterms that diverge when the regulator goes away.

There are, however, signs of missing links in that picture. Perhaps the most important one is the problem of a clean extension of the 4-dimensional helicity methods [13] to multiloop integrals where the DR+MS combo rules since the French project [14]. Another is seen in a very compact calculation of RG functions done without DR in ref. [15]. Still another sign is a mismatch between the analytical calculations with DR+MS and the numerical $g-2$ projects [16] where the subtraction procedures are still being discussed [17]. It was also discovered that DR sometimes fails in non-Euclidean situations [18], which was to be expected given that it only regulates the spacelike vector components, $3 \rightarrow D-1$, without affecting the timelike one.

It is not then surprising that 4-dimensional schemes continue to be explored [19]–[22].

Moreover, easy-to-learn prescriptions need not imply optimal execution, quite on the contrary: efficiency tends to be a result of some sophistication. Consider the analogy between the UV renormalization and the derivatives in the classical mechanics [6]. The anxieties about infinitesimals are long past, and even if finite regulators $\Delta x$ and $\Delta t$ cannot be avoided in computations, no one laments the rules of calculus to be a sweeping of dust under the carpet, nor would one forgo the power of analytical mechanics in favor of easy-to-learn finite differences on the pretext that, at some small $\Delta x$ and $\Delta t$, a new physics would kick in. Of course, it would. But when quantum mechanics arrived, the infinitesimals were not even on the list of concerns.

Our work follows the premise [3] that proper techniques for efficient work with complexly structured multidimensional integrals riddled with singularities are naturally based on the concept of generalized functions, or distributions [23] rather than the easy-to-learn subtractions. The power of generalized functions over ordinary ones has the same nature as that of real numbers over rational or complex over real. In all cases restrictions on allowed operations are lifted, opening outlandish paths to novel results via intermediacy of "imaginary" $\sqrt{2}$, $\sqrt{-1}$, the $\delta$-function, etc.

For starters, Stueckelberg et al. [24] interpreted the problem of UV renormalization as that of a proper definition of chronological products at those space-time points where some argu-ments of the singular products coalesce. Bogolyubov cast that interpretation into an accurate mathematical form of extension of functionals [25].

The elementary constructive step of such an extension by one dimension is found in proofs of the foundational Hahn-Banach theorem in any textbook of functional analysis, of which the theory of distributions is an application. (The operators $\mathbf{r}^*$ of eq. (3) are a special version of such an extension by $n$ dimensions.) A complete extension envisaged by Bogolyubov would consist of a properly arranged composition of such steps. A rigorous distribution-theoretic construction of such a composition for operator products was accomplished much later in an abstract fashion with axiomatic intention [26].

In the meantime, the famous prescription of the Bogolyubov and Parasyuk [27] was nothing but the abovementioned composition explicitly resolved (with the order of subtractions inverted compared with the order of extension steps). The answer was thus reduced to an ordinary absolutely convergent ($L^1$) integral with a bunch of so-called subtractions on top of the original "bare" integrand. Bogolyubov never explained how the prescription had come about, and physicists did not care (they rarely do) and proceeded to computations. Thus, the so-called subtraction approach was born.

However, the composition involves a branching recursion, whence one part of the difficulties of the subtraction approach, taking one all the way to Hopf algebras and Hodge structures [28], [29]. The other major obfuscation results from the use of parametric representations, where the physical multiplicative structure of the original integrals gets encrypted beyond recognition, whence smart proofs that, within DR, a $p^2$ coming from vertex factors in the numerator may indeed cancel a $p^2$ from propagators in the denominator.

Distributions were introduced into the calculations-oriented studies of multiloop integrals as a practical tool in ref. [3] where the basic mechanism of extension of functionals was augmented by another equally basic mechanism that allows the extended functional to retain the approximation properties of the non-extended one relative to a third functional, the one being approximated (the so-called consistency conditions). DR+MS provided a fast track to useful formulae (due to convenient nullifications in the so-called shadow terms, see sec. 3 below), and there quickly followed a bunch of solutions such as the R*-operation [30]; a physically correct operator product expansion for models with massless particles [31] together with efficient calculational formulae for the coefficient functions [32]; a similar result for large-$M$ effective Lagrangian and arbitrary Euclidean regimes [33], [34]. Later the techniques were extended to genuinely non-Euclidean situations [35], including unstable fields [36] and the large-$s$ regime [37].

In the meantime, the differential renormalization was proposed that exploited techniques of generalized functions [19], and Schnetz developed its "natural" variation and employed the latter for a remarkably compact calculation [15].

On the other hand, it was clear from the start that even if DR is a great facilitator, it cannot be *the* reason behind the abovementioned bunch of solutions, and that the theory of asymptotic operation could be rewritten without DR. This was accomplished by 1988, with the formalism presented in talks [4] and texts [8], [9] (the arXiv LaTex versions differ from the 1988 ChiWriter originals in format only [7]).

**3. Minimal subtractions in $D\!=\!4$ (MS$^4$ scheme).** We always consider Wick-rotated, Euclidean, $D\!=\!4$ loop integrals (both restrictions are avoidable). The MS$^4$ scheme is a fine-tuning of the GMS scheme [9]. What follows is a brief overview with emphasis on the MS$^4$ specifics, see refs. [8], [9] for further details of GMS.

Let $G$ be a one-particle irreducible (1PI) graph; $p$, the set of its loop momenta (i.e. $p\!=\!0$ means setting to zero all 4-dimensional loop momenta); $\kappa$, the entire set of $G$'s masses and external momenta; $B_G(p;\kappa)$, the $G$'s bare integrand, i.e. a product of propagators and vertex factors without subtractions or regulators. The corresponding UV renormalized integral $\mathcal{A}_G$ is defined by the GMS/MS$^4$ prescription as follows (sec. 2 of [9]):

$$\mathcal{A}_G(\kappa) \equiv \mathcal{R}_{\mathrm{UV}} \int dp\, B_G(p;\kappa) \triangleq \lim_{\Lambda\to+\infty} \int dp\, \Theta(p/\Lambda)\bigl[B_G(p;\kappa) - \mathcal{S}(p;\kappa)\bigr] \qquad (1)$$

Here $\mathcal{S}$ represents the so-called *shadow terms*. Their construction is explained in sec. 6, but a general idea is that they are exactly those terms in the asymptotic expansion of $B_G$ at $p\to\infty$ (uniformly in all directions) which are responsible for the divergence of the unsubtracted integral as $\Lambda\to\infty$, the intention being to guarantee a finite limit (1).

**4. Cutoff functions $\Theta$.** The integral on the r.h.s. of (1) contains a cutoff by means of a smooth function $\Theta(p)$ such that $\Theta(0)\!=\!1$, $\Theta'(0)\!=\!0$ and $\Theta(p)$ vanishes rapidly (i.e. faster than any power of $p$) as $p\!\to\!\infty$. This guarantees a finite limit in eq. (1) with a proper construction of the shadow terms $\mathcal{S}$. The limit value is independent of the specific shape of $\Theta(p)$ which then can be chosen to suit other purposes, e.g. to help prove equivalence of eq. (1) to the UV renormalization via quasi-local counterterms [9]. Similar cutoffs occur elsewhere in the formalism where their shape is no longer insignificant, so it is best to choose them all from the start in a coordinated, uniform fashion. Thus, the MS$^4$ fine-tunings begin with a universal choice of $\Theta(p)$ that meets all restrictions and more:

$$\Theta(p) = \prod_{g\in G} \Theta_0\bigl(p_g^2\bigr) \qquad (2)$$

The product runs over all propagators (lines) of $G$, $p_g$ is the $g$-th line's 4-momentum, and $\Theta_0(z)$ is a universal smooth function that vanishes rapidly for large $z$ and $\Theta_0(0)=1$. The choice $\Theta_0(z)=e^{-z}$ stands out for its perfect fit with the Schwinger parameters (see sec. 9). Another exponential option is $\Theta_0(z) = (1+z)\,e^{-z}$, with $\Theta_0'(0) = 0$.

A subtlety concerns the dependence of $\Theta_0\bigl(p_g^2\bigr)$ on the external momenta via $p_g$ (we call such $\Theta$'s physical). Alternatively, one can set such dependence to zero and obtain what we call pure $\Theta$'s: $\Theta_0(p_g^2) \to \Theta_0\bigl(p_g^2|_{\kappa=0}\bigr)$. The limit value of (1) is independent of whether one uses physical or pure $\Theta$'s as cutoffs. This connects our $\Theta$'s with the Polchinski cutoffs, see comment (iii) in sec. 10 below. In what follows, only pure $\Theta$'s are used.

**5. Special minimal subtraction operators $\mathbf{r}^*$.** Next, we need a special case of what mathematicians may call "regularization" (e.g. [23]), and we call "elementary subtraction operators" and generically denote as $\mathbf{r}$. (Here and below we use a heuristic language that admits a mapping into a precise reasoning in terms of test functions, their supports, etc. [8]. As in other branches of calculus, a heuristic language coupled with some educated caution is adequate for applications except for rare tricky cases.) Such operators convert (generalized) functions $F(p)$ that are non-integrable at an isolated point, say $p=0$, into completely defined distributions $\mathbf{r} \circ F(p)$. From such elementary operators, composite subtraction operations for products of singular functions are assembled in a standard recursive fashion (§11.5 of [8] where such compositions are generically denoted as $\mathbf{R}$).

The subtraction operators employed in the MS$^4$ formalism satisfy three restrictions, one is inherited from the GMS formalism, another is a fine-tuning of the finite arbitrariness, the third one is a new congruity rule explained below (and to avoid confusion we replace the tildes used in ref. [8] with asterisks). First, the definition of $\mathbf{r}^*$ is modified from the familiar Gelfand-Shilov one (eq. (3) in sec. 1.1.7 of [43]) by specializing the cutoff to the last subtracted term only (§10.8 of [8]) and fixing the remaining finite arbitrariness in a special way. Here is the formula for the one-dimensional case, straightforwardly generalized to many dimensions:

$$\int dp \left[ \mathbf{r}^*_\mu \circ F(p) \right] \varphi(p) \triangleq \int dp\, F(p) \begin{bmatrix} \varphi(p) - \varphi(0) - p\varphi'(0) - ... - \frac{1}{(n-1)!} p^{n-1} \varphi^{(n-1)}(0) \\ -\frac{1}{n!} p^n \varphi^{(n)}(0)\, \Theta(p/\mu) \end{bmatrix} + c\frac{\varphi^{(n)}(0)}{n!} \quad (3)$$

The number of subtractions $n$ is chosen to be the minimal one needed to ensure existence of the integral near zero by power counting (this is roughly similar to the absolute convergence of ordinary integrals). If $F(p)$ is integrable at zero by power counting, one sets $\mathbf{r}^* \circ F \equiv F$.

The $\mathbf{r}^*_\mu$ on the l.h.s. carries an explicit subscript to indicate the scale arbitrariness inherent in its definition and parametrized by $\mu$. Note that there is no arbitrariness with $\varphi^{(m)}(0)$, $m < n$, and that we choose $c$ in (3) to be independent of $n$ (unlike the definition (10.34) of [8] where an independent free parameter was left for each $n$).

The operators $\mathbf{r}^*$ thus defined have some handy properties:

(a) they commute with multiplication of $F(p)$ by polynomials of $p$;

(b) they preserve scaling properties of $F(p)$ (if the latter had any) to a maximal degree (i.e. modulo additive logarithmic corrections);

(c) $\int dp \left[ \mathbf{r}^*_\mu \circ F(p) \right] p^m = 0$ for $0 \leq m < n$;

(d) $\int dp \left[ \mathbf{r}^*_\mu \circ F(p) \right] p^n\, \Theta(p/\mu) = c$.

The last point concerns the $\Theta$ on the r.h.s. of (3). It is similar to the $\Theta$'s used for cutoffs in (1), so we adopt the following **congruity rule for the MS$^4$ operators $\mathbf{r}^*$**: whenever $\mathbf{r}^*$ occurs in a context similar to eq. (1) with the cutoff $\Theta(p/\Lambda)$ determined by the physical structure of $G$ according to (2), then $\mathbf{r}^*_\mu$ is defined with the cutoff $\Theta(p/\mu)|_{\kappa=0}$ (i.e. built from pure $\Theta$'s only; see sec. 4).

In sec. 7, we will explicitly need the special case of eq. (3) for $F(p)$ with a logarithmic singularity at $p \to 0$, so that $n = 0$:

$$\int dp \left[\mathbf{r}_\mu^* \circ F(p)\right] \varphi(p) = \int dp\, F(p)\left[\varphi(p) - \varphi(0)\,\Theta(p/\mu)\right] + c\varphi(0) \tag{4}$$

The option $c \neq 0$ remains useful for setting up gauge identities.

**6. Construction of the shadow terms** $\mathcal{S}$ proceeds as follows (for complete details see §3.1 of [9]):

(i) The bare integrand is asymptotically expanded for $p \to \infty$ (equivalent to $\kappa \to 0$ on dimensional grounds). To allow integrations over spherical layers, the expansion must be in the sense of distributions. Such an expansion is generated by the asymptotic operation **As** [8] that essentially is a Taylor expansion augmented by $\delta$-functional asymptotic counterterms (see an outline in sec. 7 below and ref. [8] for complete details). But in the present case it is limited to $p \neq 0$ i.e. lacks the last asymptotic counterterm located at $p = 0$. The latter omission is conventionally signaled by a prime:

$$B_G(p;\kappa) \xrightarrow{p\to\infty\ (\kappa\to 0)} \mathbf{As}' \circ B_G(p;\kappa) \tag{5}$$

All such expansions are constructed to run in powers and logarithms of the expansion parameter, therefore (§15.4 of [8]) they are defined uniquely, commute with multiplications by polynomials of $p$, and are independent of any $\mu$ that will enter the game at step (iii).

(ii) In the expansion (5), retained are exactly those terms which would generate divergencies as $\Lambda \to \infty$ after integration in (1). This restriction is indicated by the subscript 0:

$$\mathbf{As}' \circ B(p;\kappa) \to \mathbf{As}'_0 \circ B(p;\kappa); \tag{6}$$

(iii) The terms which generate specifically logarithmic divergencies at $p \to \infty$, are also logarithmically divergent near $p = 0$ (cf. $\int_0^\infty dx/x$). Therefore, a further subtraction operator $\mathbf{r}_\mu^*$ given by (4), is applied to define the distribution on the entire space including $p=0$:

$$\mathbf{As}'_0 \circ B(p;\kappa) \to \mathbf{r}_\mu^* \circ \mathbf{As}'_0 \circ B(p;\kappa) \triangleq \mathcal{S}. \tag{7}$$

This definition makes the limit (1) finite by construction.

Now the name "shadow terms" can be explained. The uniqueness of **As** (and **As**′) implies its commutativity with linear transformations of $B_G$ and with multiplications of $B_G$ by polynomials of $p$ — and the same is true for $\mathbf{r}^*$. Therefore, if $B = \sum_i \mathcal{P}_i B_i$, where $\mathcal{P}_i$ are polynomials of $p$, then $(B - \mathcal{S}) = \sum_i \mathcal{P}_i(B_i - \mathcal{S}_i)$.

Factorization of the prescription (1) for factorizing bare integrands is a less obvious implication of the MS[4] rules for $\Theta$ and $\mathbf{r}^*$: if $p = p' \oplus p''$ and $B_G(p) = B_{G'}(p') \times B_{G''}(p'')$ then $\mathcal{A}_G(\kappa) = \mathcal{A}_{G'}(\kappa) \times \mathcal{A}_{G''}(\kappa)$ (cf. Appendix A of [9]). Incidentally, this also extends the definition (1) to non-1PI graphs.

The key theoretical property of the subtraction procedure (1) thus defined is that it satisfies the sine qua non Stueckelberg-Bogolyubov causality axiom which demands that UV subtractions be effected by counterterms that are quasi-local in coordinate representation. This was shown in sec. 5 of ref. [9] for GMS schemes via a transition to DR. With the MS[4] rules for $\Theta$ and $\mathbf{r}^*$, the quasi-locality of the UV subtractions as defined in (1) is verified in a more streamlined fashion, see [2] for details.

Unlike $\mathbf{As}'$, the subtraction operators $\mathbf{r}^*_\mu$ in (4) depend on an arbitrary $\mu$. This dependence is, predictably, the source of the Callan-Symanzik equations to which we now turn.

**7. The Callan-Symanzik equation** for a single diagram is derived within MS$^4$ in a most straightforward fashion by first differentiating eq. (1) with respect to the $\mu$ in $\mathcal{S}$ (eq. (7)), and then invoking the explicit expressions for the asymptotic operation $\mathbf{As}'$ from [8], [9].

The first step is easy as all of $\mu$ sits exactly one level deep within $\mathcal{S}$. Consider the integral (1) with eq. (7) taken into account. The bare integrand $B_G$ is independent of $\mu$, and so is $\mathbf{As}'_0$ (see after (5)). The latter fact is as basic as its consequences are immediate and nontrivial. Using eq. (4), differentiating with respect to $\mu$, and dropping the $\Lambda$ limit that has become trivial, obtain a prototype CS equation:

$$\mu^2 \frac{\partial}{\partial \mu^2} \mathcal{A}_G(\kappa,\mu) + \int dp \frac{1}{2} \hat{\partial}\Theta(p/\mu) \times [\mathbf{as}'_0 \circ B_G(p;\kappa)] = 0, \tag{8}$$

where $\hat{\partial} = p_\alpha \partial_\alpha$ is the Euler operator and $\mathbf{as}'_0$ retains exactly those terms from $\mathbf{As}'_0$ which correspond to logarithmic divergence. The integral is well-defined by power counting since $\mathbf{as}'_0 \circ B_G(p;...)$ has a logarithmic singularity at $p \to 0$, whereas $\hat{\partial}\Theta(p)\big|_{p=0} = 0$.

The second step is to plug in the explicit expression for the asymptotic operation provided by the basic formula (3.11) of [9]. It is quite general and too cumbersome to quote here in full detail. However, its structure is rather regular and can be described as follows. The asymptotic operation (generically denoted as $\mathbf{As}$) is a tool to produce a correct (in the sense of distributions) asymptotic expansion of a product (such as $B_G$ above) in situations where individual factors develop non-integrable singularities after Taylor expansion (e.g. the expansion of $(p^2+m^2)^{-1}$ in $m^2$). $\mathbf{As}$ converts the product $B_G$ into a sum of a *main term* and *asymptotic counterterms*. The main term is obtained by the Taylor expansion ($\mathbf{T}$) of $B_G$, then the Taylor expansion is rendered a well-defined distribution by a term-wise application of the composite operation $\mathbf{R}_\mu$ assembled from elementary subtraction operators, in the present case $\mathbf{r}^*_\mu$ (see sec. 5): $B_G \to \mathbf{R}_\mu \circ \mathbf{T} \circ B_G$. (Note that although $\mathbf{As}$ is defined uniquely and independent of any $\mu$, its separation into the main term and the asymptotic counterterms involves an arbitrariness which we fix here according to the universal rules introduced earlier on and parametrize by the same $\mu$ as occurs elsewhere in eq. (8), which will play a role below.) However, $\mathbf{R}_\mu$ alone does not guarantee that the resulting distribution yields a correct asymptotic expansion when integrated with test functions over singularities of the Taylor expansion. To ensure that, one augments the main term $\mathbf{R}_\mu \circ \mathbf{T} \circ B_G$ by adding asymptotic counterterms localized at such singularities (the extension of functionals with the asymptotic property preserved [3]). Each such counterterm corresponds to a singular manifold of the formal expansion, and to each such manifold corresponds a so-called complete (or maximal) singular subproduct $B_\gamma$ of $B_G$. The counterterm consists of a sum of derivatives $\delta_\gamma^{(\beta)}$ of the $\delta$-functions $\delta_\gamma$ localized on that manifold with coefficients $c_{\gamma,\beta,\mu}$ constructed according to the consistency condition (for details see sec. 17 of [8]). Such counterterms replace the corresponding subproducts in $B_G$, with intact complement subproducts $\mathbf{R}_\mu \circ \mathbf{T} \circ B_{G \setminus \gamma}$. Adding all such counterterms to the main term completes the construction. Its schematic representation is this:

$$\mathbf{As} \circ B_G = \mathbf{R}_\mu \circ \mathbf{T} \circ B_G + \sum_\gamma \sum_\beta c_{\gamma,\beta,\mu} \times \delta_\gamma^{(\beta)} \times \mathbf{R}_\mu \circ \mathbf{T} \circ B_{G \setminus \gamma} \tag{9}$$

Eq. (9) is only meant to show a general structure and not all the dependencies.

All one has to do after substituting the whole thing into (8), is to integrate out those $\delta$'s explicitly, which task is trivial per se but is encumbered by the taking into account of all the accidentia such as the power-counting restriction formulated after eq. (8), the MS[4] rules for $\Theta$'s and $\mathbf{r}^*$'s, the properties of the operators $\mathbf{r}^*$, the switching of derivatives from the $\delta$'s to everything else including $\Theta$'s, various factorizations, graph combinatorics, etc. We save all the tedious accounting for a longer paper [2], and only mention the following facts that clarify the structure of the result.

First, complements $G \setminus \gamma$ for complete singular subgraphs $\gamma$ in a 1PI graph $G$ form a disjoint set of 1PI subgraphs $H$. A sum over such sets is exactly the one in (the MS variation of) the Bogolyubov R-operation, which fact is behind the physical correctness of the GMS /MS[4] subtractions (1). In the present case, however, the sum simplifies, see below.

The second fact is that the coefficients $c_{\gamma,\beta,\mu}$ have a form similar to (1) (in fact, it was a study of their structure that revealed the definition (1)). The two facts should convey an idea of how eq. (8), after using (9), becomes a per-graph Callan-Symanzik equation:

$$\mu^2 \frac{\partial}{\partial \mu^2} \mathcal{A}_G(\kappa,\mu) + \sum_H \mathcal{A}_{G/H}(\kappa,\mu) \otimes \mathcal{B} \circ H = 0 \tag{10}$$

The summation runs over all UV divergent 1PI subgraphs $H$ of $G$, including $H = G$. From eq. (9) one expects the sum to run over disjoint sets of 1PI subgraphs, but it simplifies because the $\mathcal{B}$-operator is zero on direct products of 1PI subgraphs (see sec. 8).

Further, for each $H$, $\mathcal{B} \circ H$ is a polynomial of the external parameters of $H$ (its masses and external momenta, even if some such momenta may be internal for $G$). The polynomial has degree $\omega_H$ (the usual index of UV divergence of $H$). Its coefficients are evaluated according to the rules of sec. 8 below. The subgraph $H$ in $G$ is shrunk to a single vertex (indicated by $G/H$), and the polynomial $\mathcal{B} \circ H$ is inserted as the corresponding vertex factor (the inserting is indicated by $\otimes$); the resulting diagram emerges automatically renormalized according to (1) due to the abovementioned connection of the coefficients $c_{\gamma,\beta,\mu}$ to eq. (1).

**8. The $\mathcal{B}$-operator.** Here we explain the explicit expression for $\mathcal{B} \circ H$ for renormalizable theories (all $\omega_H \leq 2$; the case of composite operators and the operator product expansions is to be treated separately) and for the simpler case of cutoffs such that $\Theta_0'(0) = 0$ (e.g. $(1+z)e^{-z}$). See ref. [2] for complete details.

Let $H$ be a 1PI divergent UV graph (regarded as a subgraph of $G$ of eq. (1)), $p_H$ the collection of its loop momenta, $\kappa_H$ its external parameters (masses and momenta; some momenta may be external for $H$ but internal for the enclosing $G$). Then:

$$\mathcal{B} \circ H \triangleq \int dp_H \, \frac{1}{2} \hat{\partial} \Theta(p_H) \times \mathbf{R}'_{\mu=1} \circ \left[ \mathbf{t}_{\omega_H} \circ B_H(p_H; \kappa_H) \right]_{\kappa_H = 0} \equiv \mathcal{P}_{\omega_H}(\kappa_H) \tag{11}$$

Note that the integral emerged from the calculation leading up to (10) with $\mu$ in the argument of $\Theta$ and in $\mathbf{R}'$ as $\ldots \hat{\partial} \Theta(p_H / \mu) \times \mathbf{R}'_\mu \circ \ldots$ This dependence vanishes after a mere change of integration variables (see below).

The algorithmic structure of (11) is as follows. One Taylor-expands the bare integrand $B_H$ in $\kappa_H$ at zero, keeping only the terms of order $\omega_H$ exactly (note that the integral, apart from the powers of $\kappa_H$, is then dimensionless, is why the above scaling works to eliminate the dependence on $\mu$). The singular result is converted into a well-defined distribution by a term-wise application of the composite operation $\mathbf{R}'_\mu$ assembled recursively according to the standard rules (§11.5 of [8]) from the MS$^4$ operators $\mathbf{r}^*_\mu$ (sec. 5). Note that $\mathbf{R}'$ is taken at $\mu = 1$ in (11), which means that $\mathcal{B} \circ H$ is a uniform polynomial of $\kappa_H$ of degree $\omega_H$, denoted as $\mathcal{P}_{\omega_H}(\kappa_H)$, with numeric coefficients given by integrals that are finite by power counting for all $p_H$ thanks to $\mathbf{R}'$. Such polynomiality in masses as well as the external momenta is a characteristic property of the entire GMS family. The case $\omega_H < 2$ requires no further comments.

For $\omega_H = 2$, one should decide how to account for the anomaly $\partial^2 p^{-2} = -4\pi^2 \delta(p)$. It is treated easiest as part of the composite $\mathbf{R} \circ \mathbf{t}$ (or $\mathbf{R} \circ \mathbf{T}$) since the separation of the asymptotic operation $\mathbf{As}$ into the main term $\mathbf{R} \circ \mathbf{T}$ and the asymptotic counterterms, as outlined above, involves just enough arbitrariness to accommodate the anomaly. But then, in order to prevent the doubling of some terms in eq. (10) compared with the prototype CS equation (8), one should demand that $\mathcal{A}_{G/H} = 0$ whenever $G/H$ degenerates into a single-propagator massless tadpole (a mnemonic: $B_{G/H}$ in this case coincides with its shadow terms $\mathcal{S}$, so $\mathcal{A}_{G/H}$ automatically nullifies according to eq. (1); it is amazing how everything fits together in this formalism, even mnemonics). Note that ref. [28] also had to deal with specifics of quadratic divergences.

An important property is that $\mathcal{B} \circ H = 0$ on factorizing graphs $H$, i.e. such that $B_H(p_H) = B_{H'}(p'_{H'}) \times B_{H''}(p''_{H''})$ with $p_H = p'_{H'} \oplus p''_{H''}$. This property explains the fact that the sum in (10) runs over single UV subgraphs (rather than disjoint sets thereof as is the case with the generic Bogolyubov R-operation). This can be ultimately traced to the Leibnitz rule for differentiating products, $(fg)' = f'g + fg'$ where the r.h.s. does not contain $f'g'$.

Now back to the assumption $\Theta_0'(0) = 0$ used to obtain eq. (11). It simplifies the switching of derivatives from the $\delta$-functions $\delta_\gamma^{(\beta)}$ in (9) to the remaining factors including $\hat{\partial} \Theta$, resulting in a term proportional to $\Theta_0'(0) \neq 0$. That term would contain an integral similar to

(11) but simpler and less singular, with $\mathbf{t}_{\omega_H} \to \mathbf{t}_{\omega_H-2}$ and with a second order monomial in $\kappa_H$ as an overall factor. An interested reader is referred to [2] for details.

**9. Complete expressions for renormalization group functions.** Eq. (10) is the Callan-Symanzik equation for a single diagram. It is extended to non-1PI diagrams by the Leibnitz rule. Then a transition to entire Green functions has to be performed. We cast eq. (10) into a form that allows a direct application of the method of ref. [34] which explicated the combinatorial details of the transition from a single diagram to the perturbative sum of all diagrams for the ultraviolet R-operation. The present case is much simpler due to the sum in (10) running over single 1PI subgraphs rather than sets of such.

The generating functional of ordinary Green functions (remember that we work with Euclidean loop integrals) is $\langle \mathrm{T} \exp(-L_{\mathrm{int}} + \varphi J)\rangle_0$, where $L_{\mathrm{int}}$ is the interaction Lagrangian. Expressions such as $\varphi^4$, $\varphi J$ etc. always imply the same space-time argument $x$ for all $\varphi$ and $J$ in the monomial, and an integration over $x$. Eq. (10) holds true for each diagram contributing to this functional. A straightforward application of the method of [34] yields:

$$\mu^2 \frac{\partial}{\partial \mu^2} \left\langle \mathrm{T} e^{-L_{\mathrm{int}}+\varphi J}\right\rangle_0 + \left\langle \mathrm{T}\left[\mathcal{B} \circ \left(\mathrm{T} e^{-L_{\mathrm{int}}}\right) \times e^{-L_{\mathrm{int}}+\varphi J}\right]\right\rangle_0 = 0 \qquad (12)$$

The $\mathcal{B}$-operator acts on all diagrams obtained by evaluating the internal T-exponent and nullifies everything but 1PI diagrams. The reasoning has been completely general so far.

Now specify $L_{\mathrm{int}} = g\varphi^4/4!$ Then $\mathcal{B}$ nullifies everything but 1PI graphs with 4 or 2 external lines, and converts such graphs into numbers in the first case and second-order polynomials of the external momentum and mass in the second. (The dimensionality is of course preserved throughout, as is Lorentz invariance.) Then one groups together various monomials via introducing the following self-evident notations for the corresponding coefficients (the latter are infinite formal series in powers of $g$ with purely numeric coefficients):

$$\mathcal{B} \circ \left(\mathrm{T} e^{-L_{\mathrm{int}}}\right) = \mathcal{B}_4 \frac{\varphi^4}{4!} + \mathcal{B}_{2,k^2} \frac{(\partial \varphi)^2}{2} + \mathcal{B}_{2,m^2} \frac{m^2 \varphi^2}{2} \qquad (13)$$

Substitute eq. (13) into (12), and after simple transformations obtain (for details see [2]):

$$\left[\mu^2 \frac{\partial}{\partial \mu^2} + \alpha(g) m^2 \frac{\partial}{\partial m^2} + \beta(g) \frac{\partial}{\partial g} + \gamma(g) J \frac{\delta}{\delta J}\right] \left\langle \mathrm{T} e^{-L_{\mathrm{int}}+\varphi J}\right\rangle_0 = 0 \qquad (14)$$

where:

$$\alpha(g) = \mathcal{B}_{2,k^2} - \mathcal{B}_{2,m^2}; \quad \beta(g) = 2g \mathcal{B}_{2,k^2} - \mathcal{B}_4; \quad \gamma(g) = \frac{1}{2} \mathcal{B}_{2,k^2} \qquad (15)$$

All the above objects are defined constructively and algorithmically.

It is straightforward to write out similar formulae for other models starting with QED.

**10. Discussion.** Contemplating the construction (11)–(15) against the prevailing multiloop practices may explain why $\mathcal{B}$ is the first letter of the word "beauty".

(i) We verified eqs. (11)–(15) by reproducing the known results (e.g. [38]) through two loops where they are independent of the renormalization scheme. We used the simplest exponential cutoff $\Theta_0(z) = e^{-z}$, and all two-loop $\mathcal{B}$-contributions reduced to two independent finite integrals, one trivial, the other simple; see [2] for details.

(ii) At this point it is not quite clear which choice for $\Theta_0(z)$ from the exponential family would lead to simpler multiloop calculations, $e^{-z}$ or $(1+z)e^{-z}$. The single integral in eq. (11) with the second option comes at the cost of proliferation of terms in the integrand due to the factor $(1+z)$ per each line; but the extra terms are simpler than the primary integrand (each $z$ cancels a propagator and simplifies the $\mathbf{R}'$), and the issue is even less clear for gauge models where one deals with large polynomials in the numerators anyway.

(iii) A fact to ponder is that the MS$^4$ rule for cutoffs (2) is equivalent to the modification, due to Polchinski [12], of the free Lagrangian within the general non-perturbative theory of renormalization group, whereas in the MS$^4$ formalism, it plays a significant role in the perturbative derivation of the differential renormalization group equation.

(iv) The expression $\left[\mathbf{t}_{\omega_H} \circ B_H\left(p_H; \kappa_H\right)\right]_{\kappa_H = 0}$ in (11) (it is a part of the formal Taylor expansion in all masses and external momenta of $B_H$) represents a maximal simplification of the original bare integrand in regard of its dependencies on dimensional parameters, and is scale invariant under $p_H \to \lambda p_H$. The scale invariance is broken by $\mathbf{R}'$ (logarithmic corrections come from the elementary operators $\mathbf{r}^*$) and by $\Theta_0$. Cf. also comment (ix).

(v) With the exponential cutoffs, the integrals generated by the $\mathcal{B}$-operator for any diagram are algorithmically reduced, via a transition to the Schwinger parameters, to $L^1$ integrals of rational functions over hypercubes. This proposition is completely general and remains valid for non-scalar models starting with QED.

(vi) An equation similar to (10) is found in a stupendous treatise by Brown and Kreimer [28] (see their Theorem 61; we would not recognize it amidst their mathematical mayhem if it were not explicitly marked as a "renormalization group equation"). An obvious difference of our formalism from theirs is that in our case the explicit expression for $\mathcal{B} \circ H$ is an immediate result of the derivation of (10). Ref. [28] reduces evaluation of RG functions to what they call single-scale integrals (cf. comment (iv)) that are also algorithmically reduced to $L^1$ integrals of rational functions over hypercubes [39], cf. comment (v). So, the two wildly different formalisms lead to similar integrals, although the exact relation between them is not clear at the moment.

Note that the Hopf algebra formalism of [28] serves the same purpose as the distribution-theoretic method of asymptotic operation [8], namely, to tame the branching compositions of subtractions. The difference is that the method of asymptotic operation postpones resolving subtraction operators to the last, managing to do all theoretical work directly with the distributions that pack the unresolved compositions, whereas ref. [28] starts out by resolving all the subtractions in the manner pioneered by Bogolyubov and Parasyuk, and then carries the resulting complexity throughout proofs and derivations. Those familiar with computer programming would note that the difference is similar to the one between a functional program that implements a branching recursion, and an equivalent purely imperative one with all the recursions expanded into nested loops.

(vii) It is theoretically possible to choose $\Theta_0(z)$ as a step function: if $z<1$ then 1 else 0, resulting in a somewhat messy but interesting integral over a hypersurface in (11).

(viii) An interesting task is numerical computation of eq. (1) which can always be algorithmically resolved into an absolutely convergent form in momentum representation. A related task is to devise an efficient way to generate parametric representations with $L^1$ integrands for eq. (1). Remember that within the distribution-theoretic approach, one resolves the composite subtraction operations after all theoretical issues (convergence, gauge symmetries, etc.) are settled, as the very last step prior to actual integrations, numerical or analytical (cf. the above expressions for RG functions). This means that the form of the resolved $L^1$ integrands need not be amenable to theoretical manipulation but only feedable into integration routines.

(ix) The $\mathcal{B}$-operator involves a nullification of *all* masses and external momenta in the bare integrand, which is a holy grail of the so-called infrared rearrangement methods starting from purely algebraic tricks [40] to the R*-operation [30] (that R* is not to be confused with compositions **R** of the MS$^4$ operators $\mathbf{r}^*$ that occur e.g. in the reasoning of sec. 7). The role of the remaining nonzero dimensional parameter is taken over by the cutoff function $\Theta_0(z)$. The $\mathcal{B}$-operator can therefore be regarded as an ultimate form of the R*-operation, with the universal structure of IR subtractions known and implementable in advance, once and for all, which makes the old R*-operation vanish, in a sense. This appears to agree with the "no R*-operation" claim of [39], given that ref. [28] arrives at similar integrals (cf. comment (vi)).

(x) The ultimate simplification of the bare integrand by the $\mathcal{B}$-operator means a maximal imaginable reduction of the variety of resulting integrals — a most welcome feature for calculations in non-abelian gauge models. The role of breaking the scale-invariance is passed to the exponential cutoffs whose structure follows the topology of the diagram and which morph into a mere change of the integration limits in the parametric representation.

(xi) In order to manipulate the renormalized integrals within the MS$^4$ scheme, one generally needs a few simple rules (commutativity with polynomials etc., see sec. 6 about the name "shadow terms"). This is actually similar to the DR where an accurate definition is cumbersome but one gets by in practice with a few simple recipes. A conclusion suggests itself that the MS$^4$ pair of rules — for the cutoffs $\Theta$ (sec. 4) and for the operators $\mathbf{r}^*$ (sec. 5) — essentially demystifies the DR+MS magic.

(xii) Regarding applicability of the MS$^4$ formalism to gauge theories, it would be sufficient to verify the Ward-Takahashi-Slavnov-Taylor identities for the amplitudes defined by (1). As comment (xi) implies, there really is no magic behind DR+MS, so, if gauge invariance is achieved with the DR+MS combo for a song, by a mere $4 \to D$, then MS$^4$ would handle it without significant complications as well. Leaving a detailed discussion to a separate paper, we only make here a few preliminary notes: (a) a key role in per-diagram proofs of gauge identities is played by invariance of integrands with respect to translations of integration momenta; (b) such invariance could only be broken within MS$^4$ by surface terms resulting from translations of the shadow terms; (c) for logarithmically divergent $B_G$ in eq. (1), such surface terms are zero, whereas for, say, linearly divergent self-energies they are irrelevant for reasons of covariance; (d) the triangle anomaly (essentially a surface term too) is handled within GMS/MS$^4$ most naturally [2].

**11. Conclusions.** The MS$^4$ scheme introduced in this Letter is natural throughout: it is a fine-tuning of the GMS scheme by the physically natural Polchinski cutoffs (sec. 4), its congruent elementary subtraction operators are singled out by handy properties (sec. 5), and the GMS renormalized integrals first emerged as coefficients of asymptotic counterterms of the asymptotic operation, thus forming a tight recursive pattern. As to the method of asymptotic operation, it emerged, again, most naturally from examination of the mathematical nature of asymptotic expansions of multiloop integrals [5]. Even in its earliest variant [3] that hooked up to the readily available dimensional regularization for quick results, the method yielded an array of novel calculational tools (listed in sec. 2), which was exactly why a focused effort was spent [8], [9] on studying it independently of extraneous regularizations on top of the cutoffs inherent in any definition of integrals over infinite momentum space. It is unfortunate that that line of research was interrupted by causes that are alien to the scientific endeavor [7].

It now transpires that the GMS/MS$^4$ scheme constitutes a fully operational universal framework. Although perturbative calculations are a vast field and it is impossible for a novel formalism to cover all the bases at once, its first specific result — the explicit formulae for renormalization group functions (secs. 8, 9) — is such that it makes one of us (FT) wish it were available back in 1979 [41] as the $\mathcal{B}$-operator technique is directly applicable to that kind of calculations. Moreover, none of the applications we have examined so far [1] revealed a serious obstacle for the new formalism, the caveat being that one must learn the techniques of distributions to apply it properly (for elementary introductions see [42], [43]).

On the theoretical side, an issue to finalize is that of gauge identities in the MS$^4$ formalism; as comment (xii) in sec. 10 implies, the issue is of a routine nature rather than a challenge. Note also that the GMS formalism was designed from the start for proofs of operator product expansions etc., so there are no problems in that quarter either.

On the practical side, the interesting tasks would be: 1) to obtain an absolutely integrable parametric representation for (1) with the $g-2$ applications in view [16] since the MS$^4$ scheme disentangles the UV and IR divergencies better than the on-mass-shell subtraction scheme, with potential benefits for the difficult numerics; 2) to adapt the basic integration-by-parts reduction algorithms [44] to the MS$^4$ formalism, having in view calculations of OPE coefficient functions via the algorithms of [32] (cf. the discussion in [1]); 3) to adapt Panzer's algorithms for hyperlogarithms [45] or, perhaps, the 'perturbative quantum geometries' (for a recent review see [46]), to the integrals generated by the $\mathcal{B}$-operator.

Finally, the strategic goal remains a marriage of the 4-dimensional MS$^4$ formalism for loop integrals to the helicity formalism [13], [20], [47] that proved so efficient at the tree level.

**Acknowledgments.** We thank Aleksey Lokhov for bringing together the team, Sergey Volkov for discussions and references, and Rinat Menyashev for supporting this work.